\documentclass[slac_one]{revtex4}
\usepackage{graphicx}
\usepackage{fancyhdr}
\renewcommand{\headrulewidth}{0pt}
\renewcommand{\footrulewidth}{0pt}

\newcommand{\text}{\rm }

\newcommand{\gag}{\ensuremath{g_{{\rm a}\gamma}}}
\newcommand{\ma}{\ensuremath{{m_{\rm a}}}}

\begin{document}

\title{Solar axion search with the CAST experiment} 

%


\author{S. Aune, E. Ferrer-Ribas\footnote{Corresponding author},Y. Giomataris, T. Papaevangelou}
\affiliation{IRFU, Centre d'\'Etudes Nucl\'eaires de Saclay (CEA-Saclay), Gif-sur-Yvette, France}

\author{D. Autiero\footnote{Present address: Institute de Physique
Nucl\'eaire, Lyon, France}, K. Barth, S. Borghi\footnote{Present address: Department of Physics and Astronomy, University of Glasgow, Glasgow, UK}, M. Davenport, L. Di Lella\footnote{Present address:
Scuola Normale Superiore, Pisa, Italy}, N. Elias, C. Lasseur, T.~Niinikoski, A.~Placci, H.~Riege, L.~Steward, L.~Walckiers, K. Zioutas}
\affiliation{European Organization for Nuclear Research (CERN), Gen\`eve, Switzerland}

\author{D. H. H. Hoffmann, M. Kuster, A. Nordt}
\affiliation{Technische Universit\"{a}t Darmstadt, IKP, Darmstadt, Germany}

\author{ H.~Br\"auninger, M. Kuster, A. Nordt}
\affiliation{Max-Planck-Institut f\"{u}r extraterrestrische Physik, Garching, Germany}

\author{B.~Beltr\'an\footnote{Present address: Department of Physics,
Queen's University, Kingston, Ontario}, J. M. Carmona, T. Dafni, J.~Gal\'an, H. G\'omez, I. G. Irastorza, G. Luz\'on, J. Morales, A. Ortiz, A.~Rodr\'iguez, J. Ruz, J. Villar}
\affiliation{Instituto de F\'{\i}sica Nuclear y Altas Energ\'{\i}as, Universidad de Zaragoza, Zaragoza, Spain}

\author{J. I. Collar, D. Miller}
\affiliation{Enrico Fermi Institute and KICP, University of Chicago, Chicago, IL, USA}

\author{C.~Eleftheriadis, A. Liolios, E. Savvidis}
\affiliation{Aristotle University of Thessaloniki, Thessaloniki, Greece}

\author{G. Fanourakis, K. Kousouris\footnote{Present address: Fermi National Accelerator Laboratory, Batavia, Illinois, USA}, T. Geralis, }
\affiliation{National Center for Scientific Research ``Demokritos'', Athens, Greece}

\author{J. Franz, H. Fischer, F. H. Heinsius,D. Kang\footnote{Present address: Institut f\"ur Experimentelle Kernphysik, Universit\"at Karlsruhe, Karlsruhe, Germany},  K.~K\"onigsmann, J. Vogel}
\affiliation{Albert-Ludwigs-Universit\"{a}t Freiburg, Freiburg, Germany}

\author{A. Belov, S.~Gninenko}
\affiliation{Institute for Nuclear Research (INR), Russian Academy of Sciences, Moscow, Russia}

\author{M. Hasinoff}
\affiliation{Department of Physics and Astronomy, University of British Columbia, Department of  Physics, Vancouver, Canada}

\author{J. Jacoby}
\affiliation{Johann Wolfgang Goethe-Universit\"at, Institut f\"ur Angewandte Physik, Frankfurt am Main, Germany}

\author{R. Kotthaus, G. Lutz, G. Raffelt, P. Serpico\footnote{Present address: European Organization for Nuclear Research (CERN), Gen\`eve, Switzerland}}
\affiliation{Max-Planck-Institut f\"{u}r Physik, Munich, Germany}

\author{K.~Jakov\v{c}i\'{c}, M.~Kr\v{c}mar, B.~Laki\'{c}, A.~Ljubi\v{c}i\'{c}}
\affiliation{Rudjer Bo\v{s}kovi\'{c} Institute, Zagreb, Croatia}

\author{Y. Semertzidis, M. Tsagri, K. Zioutas}
\affiliation{Physics Department, University of Patras, Patras, Greece}

\author{K. van Bibber, M. J. Pivovaroff, R. Souffli}
\affiliation{Lawrence Livermore National Laboratory, Livermore, CA, USA}

\author{E. Arik$^1$\footnote{1: Deceased}, F. S. Boydag$^1$, S. A. Cetin, O. B. Dogan$^1$, I. Hikmet$^1$}
\affiliation{Dogus University, Istanbul, Turkey}

\author{G. Cantatore, M. Karuza, V. Lozza, G. Raiteri}
\affiliation{Instituto Nazionale di Fisica Nucleare (INFN), Sezione di
Trieste and Universit\`a di Trieste, Trieste, Italy}

\author{S. K. Solanki}
\affiliation{Max-Planck-Institut f\"{u}r Aeronomie, Katlenburg-Lindau, Germany}

\author{T. Karageorgopoulou, E. Gazis}
\affiliation{National Technical University of Athens, Athens, Greece}

\begin{abstract}
The CAST (CERN Axion Solar Telescope) experiment is searching for solar axions by their conversion into photons inside the magnet pipe of an LHC dipole.
The analysis of the data recorded during the first phase of the experiment with vacuum in the magnet pipes has resulted in the most restrictive experimental limit on the coupling constant of axions to photons.
In the second phase, CAST is operating with a buffer gas inside the magnet pipes in order to extent the sensitivity of the experiment to higher axion masses.  We will present the first results on the $^{4}{\rm He}$ data taking as well as the system upgrades that have been operated in the last year in order to adapt the experiment for the $^{3}{\rm He}$ data taking. Expected sensitivities on the coupling constant of axions to photons will be given for the recent $^{3}{\rm He}$ run just started in March 2008.
\end{abstract}

\maketitle

\thispagestyle{fancy}

\section{INTRODUCTION} 

The CAST (Cern Axion Solar Telescope) experiment is using a
decommissioned LHC dipole magnet to convert solar axions into
detectable x-ray photons. Axions are light pseudoscalar particles
that arise in the context of the Peccei-Quinn\cite{PecceiQuinn}
solution to the strong CP problem and can be Dark Matter
candidates\cite{Sikivie}. Stars could produce axions via the
Primakoff conversion of the plasma photons. The CAST experiment
is pointing at our closest star, the Sun, aiming to detect solar axions. The
detection principle is based on the coupling of an incoming axion
to a virtual photon provided by the transverse field of an intense
dipole magnet, being transformed into a real, detectable photon
that carries the energy and the momentum of the original axion.
The axion to photon conversion probability is proportional to the
square of the transverse field of the magnet and to the active
length of the magnet. Using an LHC magnet ($9\;$T and $9.26\;$m long)
improves the sensitivity by a factor 100 compared to previous
experiments. 
The CAST experiment has been taking data since 2003 providing the most restrictive
limits on the axion-photon coupling~\cite{Zio05,And07} for masses 
$\ma\lesssim 0.02\;$eV. At this mass the sensitivity is degraded due to coherence loss. 
In order to restore coherence, the magnet can be filled with a buffer 
gas providing an effective mass to the photon\cite{vanBibber:1988ge}. By changing the pressure of the buffer gas in 
steps, one can scan an entire range of axion mass values.  At the end of 2005 the CAST 
experiment started such a program, entering its phase II by filling the magnet bore with He gas.
From 2005 to 2007, the magnet bore was filled with $^{4}{\rm He}$ gas extending our sensitivity to masses up to 
$0.4\;$eV, preliminary results will be presented here. From March 2008 onwards the magnet bore has been filled with $^{3}{\rm He}$ and  
the sensitivity should be increased to sensivities up to $\ma\lesssim1.2\;$eV by the end of the $^{3}{\rm He}$ run in 2010. 

\section{THE CAST EXPERIMENTAL SET UP: RECENT UPGRADES}
The CAST set up has been described elsewhere~\cite{Zio05,Zio99}. From 2002 to 2006
three X-ray detectors were mounted on the two sides of the magnet: a conventional TPC\cite{Aut07} 
covering both magnet bores looking for sunset axions; in the sunrise side one of the bores was covered by a Micromegas detector\cite{Abb07} and in the other bore a CCD detector coupled to a telescope\cite{Kus07} improving the signal to background ratio by a factor 150. In 2006 the TPC started to show a degraded performance due to aging. It was then decided to replace the sunset TPC and the existing Micromegas detector in the sunrise side by a new generation of Micromegas detectors\cite{Bulk} that coupled with suitable shielding would improve greatly their performance. The new detectors
were commissioning end of 2007 and by mid 2008 they have already shown an improvement in performance that has been translated in a background reduction of a factor 15 compared to the TPC performances and a factor 3 compared to the standard Micromegas detector used without shielding till 2006.
In 2005, the experiment went through a major upgrade to allow operation with He buffer gas in the cold bore. This upgrade was done in two steps: first the system was designed for operation using $^{4}{\rm He}$ and in 2007 the system was upgraded for operation at higher buffer gas densities using $^{3}{\rm He}$. The system has been designed to control the injection of He in the magnet bores with precision and to monitor accurately the gas pressure and temperature\cite{Tapio08,He3TDR}. Special care has been taken to achieve high precision in the reproducibility of steps ($< 0.01\;$mbar) and to protect the system for $^{3}{\rm He}$ loss. The $^{3}{\rm He}$ system has been operating succesfully since december 2007.

\section{RESULTS}
As during phase I, the tracking data (magnet pointing the sun) represented about 2$\times$1.5 hours per day while the rest of the day was used to measure background.  The procedure was to daily increase the $^{4}{\rm He}$ density so that sunrise and sunset detectors measure every pressure. Every specific pressure of the gas allows to test a specific axion mass having a new discovery potential. The $^{4}{\rm He}$ data recorded  end of 2005 and 2006 represents around 300 hours of tracking data and 10 times more hours of background data for each detector, covering 160 pressure settings allowing to scan a new axion mass range between 0.02 and 0.39~eV.

\begin{figure*}[t]
\centering
\includegraphics[width=8cm]{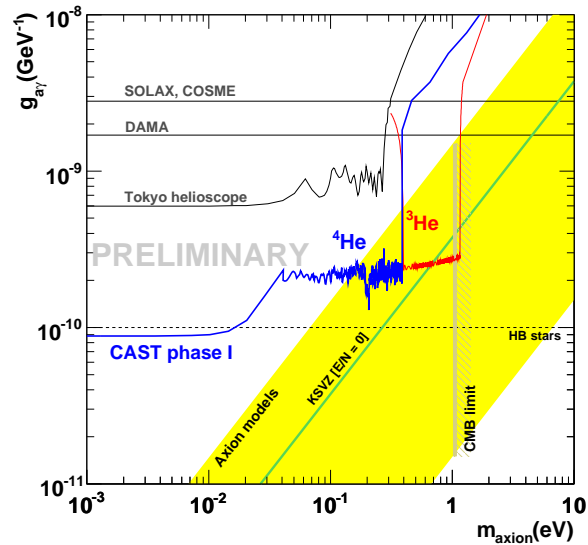}
\caption{Exclusion plot in the axion-photon coupling versus  axion
   mass plane. The limit achieved by the CAST experiment (combined result of the CAST phase I
   and $^{4}{\rm He}$ part of phase II) is compared with constraints obtained from the Tokyo helioscope and HB stars. The red dashed line shows our prospects for the $^{3}{\rm He}$ run started in March 2008. The vertical line (HDM) is the
   hot dark matter limit for hadronic axions
   $\ma<1.0~{\rm eV}$ inferred from observations of the cosmological large-scale structure.
   The yellow band represents typical theoretical models with
   $\left|E/N-1.95\right|$ in the range 0.07--7 where the green solid line
   corresponds to the case when $E/N=0$ is assumed.} \label{CASTexclusion}
\end{figure*}

\begin{figure*}[t]
\centering
\includegraphics[width=9cm]{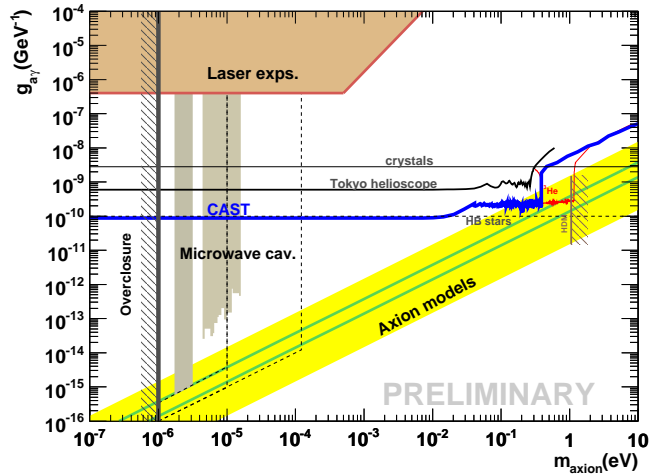}
\caption{Exclusion plot in the axion-photon coupling versus the axion mass plane for a wider range of parameters. Limits from laser, microwave and crystal axion searches have been included.} \label{Exclusion}
\end{figure*}

An independent analysis was performed for each data set of the three different detectors. A combined preliminary result was derived  where from the absence of a signal above background CAST excludes a new range in the \gag--\ma~plane shown in figure~\ref{CASTexclusion} from axion masses of 0.02\,eV (Phase I) up to masses of 0.39\,eV. This parameter space was not previously explored in laboratory experiments. CAST has therefore entered the QCD axion band for the first time in this range of axion masses, excluding an important portion of the axion parameter space. Figure~\ref{Exclusion} shows the exclusion plot for a wider range of masses including limits obtained by other type of searches (laser, microwave and crystals). The final results will be published soon in \cite{He4CAST}.

\section{CONCLUSIONS}
The CAST experiment has established the most stringent experimental limit on axion coupling constant over a wide range of masses, exceeding astrophysical constraints. The $^{4}{\rm He}$ phase has allowed to enter in an
unexplored region favoured by the theory axion models. From the absence of excess X-rays when the magnet was
pointing to the Sun, we set a preliminary upper limit on the axion-photon coupling of $\gag\lesssim 2.22\times
10^{-10}\,{\rm GeV}^{-1}$ at 95\% CL for $\ma \lesssim 0.4$~eV, the exact result depending on the pressure setting. At present, with the $^{3}{\rm He}$ run we are exploring deeper this region to reach sensitivities of $\ma<$1.2\,eV.


\end{document}